\documentclass[mathleft
]{an}
\usepackage{graphicx}
\usepackage{times}

\def\pg{{PG\,1211+143}}

\def\le{{L_{\rm Edd}}}

\def\me{{\dot M_{\rm Edd}}}

\def\mw{{\dot M_{\rm w}}}
\def\red{{R_{\rm Edd}}}

\def\xmm{{\it XMM-Newton}}
\def\chandra{{\it Chandra}}
\def\suzaku{{\it Suzaku}}
\def\et{{et al.\ }}

\def\hitomi{{\it HITOMI}}
\def\athena{{\it Athena}}

\newcommand{\ls}{\mathrel{\hbox{\rlap{\hbox{\lower4pt\hbox{$\sim$}}}\hbox{$<$}}}}
\newcommand{\gs}{\mathrel{\hbox{\rlap{\hbox{\lower4pt\hbox{$\sim$}}}\hbox{$>$}}}}

\def\Msun{\hbox{$\rm ~M_{\odot}$}}

\def\H0{{\rm ~km~s^{-1}~Mpc^{-1}}}

\def\et{{et al.}}

\begin{document}

% The following seven commands are intended for editorial usage and should be
%ignored by
% the author(s).
\Pagespan{1}{}% Document's page range. 
% If second parameter is left empty, the last page is computed automatically.
\Yearpublication{2017}%
\Yearsubmission{2016}%
\Month{??}%   
\Volume{???}%  
\Issue{??}% 
\DOI{??}% 
\date{\today}

\title{Exploring accretion disc physics and black hole growth with regular monitoring of ultrafast AGN winds}

\author{Ken Pounds, Andrew Lobban and Chris Nixon}

\titlerunning{Powerful AGN winds}
\authorrunning{K.A.Pounds \et}
\institute{
  Department of Physics and Astronomy, University of Leicester, Leicester LE1 7RH
  UK}

\received{September 2016}
\accepted{December 2016}
\publonline{}


\abstract{15 years of \xmm\ observations have established that
  ultra-fast, highly ionized winds (UFOs) are common in radio-quiet
  AGN. A simple theory of Eddington-limited accretion correctly
  predicts the typical velocity ($\sim$0.1$c$) and high ionization of
  such winds, with observed flow energy capable of ejecting
  star-forming gas. An extended \xmm\ observation of the archetypal UFO,
  \pg\, recently found a more complex flow pattern, suggesting
  that intensive \xmm\ observations offer exciting
  potential for probing the inner accretion disc structure and
  SMBH growth.}

\maketitle

\section{Introduction}

X-ray spectra from an \xmm\ observation of the luminous Seyfert galaxy
\pg\ in 2001 provided the first detection of strongly blue-shifted
absorption lines of highly ionized gas in a non-BAL Active
Galactic Nucleus (AGN), corresponding to a sub-relativistic outflow
velocity of $\sim$0.1$c$ (Pounds \et\ 2003), later adjusted to
0.15$\pm$0.01$c$ with the identification of additional absorption lines
and broad-band spectral modelling (Pounds \& Page 2006). Further
observations of \pg\ over several years with \xmm, \chandra\ and
\suzaku\ found the high velocity outflow to be persistent but of
variable strength (Reeves \et\ 2008). Evidence that the outflow in
\pg\ was both massive and energetic - with potential importance for
galaxy feedback (King 2003, 2005) - came from the detection of a P-Cygni line
profile and other broad emission features by combining the 2001, 2004
and 2007 \xmm\ EPIC spectra (Pounds \& Reeves 2009).
Examination of archival data from \xmm\ and \suzaku\ has since shown
ultra-fast, highly-ionized outflows (UFOs) to be common in
nearby, luminous AGN (Tombesi \et\ 2010, 2011; Gofford \et\ 2013).
The frequency of these detections suggest a wide angle outflow, with mass rate and kinetic energy in a persistent wind
capable of curtailing star formation and black hole growth
(Pounds 2014), and providing the 
link between black hole and stellar bulge masses implied by the
observed $M-\sigma$ relationship (Gebhardt \et\ 2000; Ferrarese \& Merritt 2000).

An extended \xmm\ observation of \pg\ in 2014 has now provided uniquely high quality X-ray spectra of \pg, revealing previously unseen velocity structure in the $\sim$6--10\,keV energy band. Spectral modelling of data from the high-energy pn camera (Strueder \et 2001) identified the observed absorption lines with resonance and higher order transitions of highly-ionized Fe, consistent with two distinct outflow velocities $v \sim 0.066c$ and $v \sim 0.13c$ (Pounds \et\ 2016a: hereafter P16).

In that first report of a dual velocity, primary (high column) wind it
was suggested that the simultaneous observation of two `Eddington
Winds' (King \& Pounds 2015) might be evidence of `chaotic
accretion', where a clump of matter approaching the black hole divides
to form prograde and retrograde accreting flows, with a corresponding
difference in accretion efficiency and potential wind
velocity. Marginal evidence in P16 for a third velocity component component ($v \sim 0.19c$)
has since been supported
in an analysis of the higher resolution soft X-ray spectra from the
Reflection Grating Spectrometer (RGS; den Herder \et\ 2001) during the same 2014 \xmm\
observation (Pounds \et\ 2016b).   Here we discuss that further challenge
to the simple Eddington Wind model and 
suggest that future well-sampled
observations of a powerful UFO, as observed in \pg, have exciting
potential to study the accretion physics determining how the super-massive
black hole (SMBH) continues to grow in the interval between mergers.

\section{Eddington Winds}

The `Black Hole Winds' model (King \& Pounds 2003) provides a
simple physical basis for high velocity outflows in AGN accreting at
or above the Eddington limit, and has the bonus of useful predictive
power.

Based on the initial observation of \pg\ the model assumes a primary (high column) wind will have electron
scattering optical depth $\tau \sim 1$,
measured inwards from infinity to a distance of order the
Schwarzschild radius. Since on average every
photon emitted by the AGN scatters about once before escaping to
infinity, and gives up all its momentum to the wind, the
total (scalar) wind momentum will be of order the photon momentum. Equating the wind and photon
momenta then gives:
\begin{equation} \dot{M}_{\rm w} v \simeq {L_{\rm Edd}\over c},
\label{mom}
\end{equation}

where $\mw$ and $v$ are the mass rate and velocity in the wind and $\le$ is the Eddington luminosity.
For accretion from a disc, as here, the classic
paper of Shakura \& Sunyaev (1973) finds a similar result at
super--Eddington mass inflow rates, with the excess accretion being expelled
from the disc in a quasi-spherical wind.
Since the Eddington accretion rate is:
\begin{equation}
\me = {\le \over \eta c^2},
\label{edd}
\end{equation}

the Eddington wind terminal velocity is then: 
\begin{equation}
v \simeq {\eta\over \dot m}c \sim 0.1c,
\label{v}
\end{equation}

where $\dot m$ is the accretion ratio, typically of order unity in SMBH growth episodes, as relevant here (see
Section 3 of King
\& Pounds 2015 for a further discussion).

Assuming the wind is launched and then coasts, the observed outflow
velocity will be of order the escape velocity at the launch
radius, providing a potential probe of the accretion geometry from observation of the wind profile.
For \pg, wind velocity components of  $v\sim 0.13c$ and $v\sim 0.066c$ correspond to
$R_{\rm launch} \sim 60R_{\rm s}$ and $\sim 240R_{\rm s}$, respectively. For a SMBH mass of $3 \times 10^{7} \Msun$
(Kaspi \et\ 2000) and
$R_{\rm s} = 2GM/c^2$, the respective launch radii are then
$\sim 4\times10^{14}$~cm and $\sim 1.6\times10^{15}$~cm.

From King \& Pounds (2015) we also note the ionization parameter of the wind:
\begin{equation} 
\xi = {L_i\over NR^2}
\label{ion}
\end{equation}
essentially determines which spectral lines are observed.
Here $L_i = l_iL_{\rm Edd}$ is the ionizing luminosity, with $l_i< 1$ specified by the AGN spectrum, and $N = \rho/\mu m_p$ 
the number density of the UFO gas.  Assuming mass conservation in the wind the ionization
parameter is then:

\begin{equation}
\xi = 3\times 10^4\eta_{0.1}^2l_2\dot m^{-2} = 3\times 10^4v_{0.1}^2l_2
\label{ion2}
\end{equation}
where $l_2 = l_i/10^{-2}$, and $\eta_{0.1} = \eta/0.1$. 

Powerful AGN winds are therefore most likely to be observable in the
X-ray spectrum, and by current instruments such as the pn and MOS
cameras on \xmm

\begin{figure}                                                          
\centering                                                              
\includegraphics[width=7.5cm, angle=0]{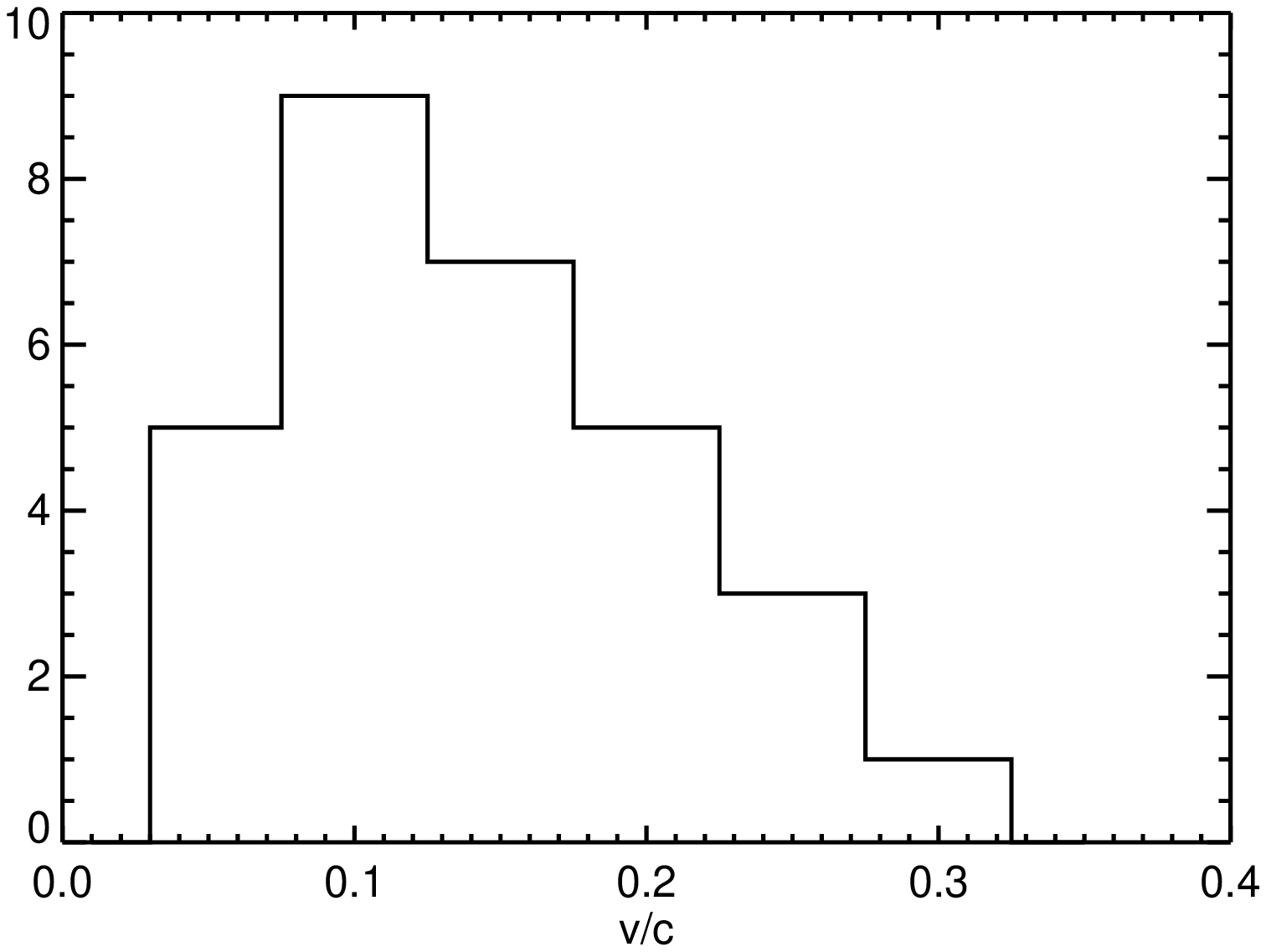}
\centering
\includegraphics[width=7.5cm, angle=0]{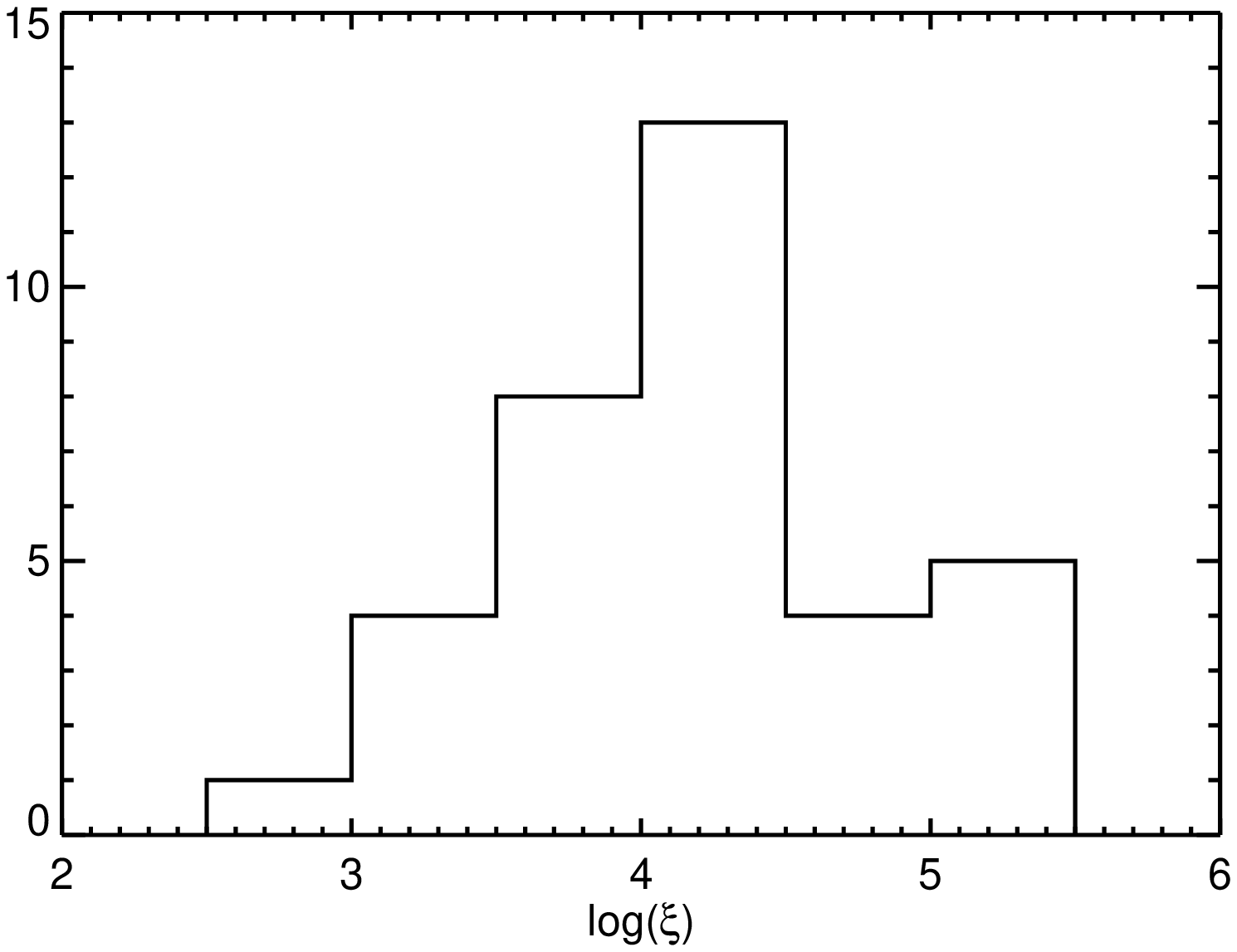}   
\caption
{Distribution of primary outflow velocity and ionization parameter in 35 radio-quiet
 AGN with UFOs detected in the \xmm\ or \suzaku\ archival surveys. The observed properties are
 consistent with the
 predictions of an Eddington wind.}
\end{figure}

Figure 1 summarises data from the 35 best-determined UFOs from the \xmm\ and \suzaku\ archival searches,
showing a distribution
of wind velocity and ionization parameter consistent with the above predictions of an Eddington wind.
We suggest in the remainder of this paper that by using the wind velocity as a marker of the region in
the disc where the accretion rate first exceeds the Eddington rate, well sampled observations of UFOs over the
next decade of \xmm\  could provide direct observational constraints on the detailed way in which a SMBH grows
between mergers.

\subsection{The origin of a unique wind velocity}

As noted above, Shakura \& Sunyaev (1973) first
pointed out that a black hole supplied with matter at a
super-Eddington rate will expel matter from its accretion disc so as
to never exceed the local Eddington luminosity.

In such a situation, which should be relevant to a luminous AGN such as \pg,  where a comparison of bolometric
luminosity and black hole mass indicates a mean accretion rate close to the Eddington rate, it is interesting to
follow a clump of additional matter  moving inwards on the local viscous timescale.

While the disc successfully radiates accretion energy at large disc radii and there is no wind, a critical point
arises
where the local accretion luminosity reaches the Eddington rate. This defines the Eddington radius ($\red$), where
\begin {equation}
\red = {\dot m GM \over \le}.
\label{red}
\end{equation}

The wind starts at $\red$, with
the correct (local escape) velocity. Mass loss from within this radius falls quickly, since
the local accretion rate $\dot M(R)$ will scale so as to keep $GM\dot M(R)/R$  $\simeq$ $\le$,
and $\dot M(R)$ $\propto$ R. Hence there
is less absorption from smaller radii which may show as a higher
velocity wing to the primary outflow.
Figure 2 illustrates the inflow of such an incident of enhanced accretion.

\begin{figure*}                                                          
\centering                                                              
\includegraphics[width=12cm, angle=0]{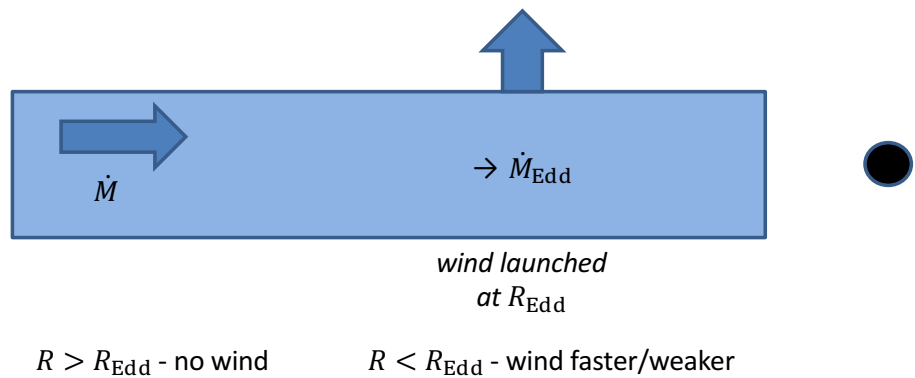}                     
\caption{Cartoon depicting the inflow of an incident of enhanced accretion. On reaching
  a critical radius in the disc where the accretion rate exceeds the local Eddington rate a fast wind is
  launched, as described in the text. While there is no wind at larger radii, inwards of
 $\red$  a weaker, higher velocity wind might be seen.}
\end{figure*}

Observation of primary (high column) wind components with
different UFO velocities in \pg\ suggests that this simple picture of a flat
axisymmetric disc does not always apply. As noted in Section 1, P16 reported the detection of two simultaneous
primary outflow velocities and speculated that the dual velocities
identified in the high quality Fe\,K absorption spectrum might be explained by `chaotic
accretion' (King \& Pringle 2006), consisting of  many prograde and retrograde events. The former
would remain stable to smaller radii, thereby releasing more gravitational energy, a higher accretion efficiency,
$\eta$, and
a higher launch velocity than the retrograde flow, for $\dot m$ $\sim$1.

It is interesting to note a recent study of accreting filaments close to Sgr A$^\star$ showing
how an accreting cloud might divide on approach to the black hole (Lucas \et\ 2013), with the potential to produce
prograde and retrograde accretion flows.
While X-ray monitoring of AGN winds is at present too limited to show how common such dual
velocities winds might be. Meanwhile, a still  more complex picture is emerging, with the recent
confirmation of a third high velocity component in the primary wind from \pg.

In Section 3 we review the current observational evidence; then in Section 4 we briefly consider how such a multiple
velocity 
primary (high-column) wind might result where an accreting `cloud' approaches the
disc at an oblique angle to the
black hole spin plane, causing the inner disc to warp and break up. We suggest such a `breaking disc'
(Nixon \et\ 2012) could launch multiple winds when fragments of the warped inner disc
collide and cool, with
the possibility of near-radial infall creating new regions of
super-critical accretion. Since the pattern of wind velocities
corresponds to the geometry of such accretion structure, we suggest that
future well-sampled observations of powerful AGN winds could provide
unique information on the complexities of SMBH accretion in the
lengthy interval between mergers.

\section{Complex velocity structure in the powerful ionized wind of
  the luminous narrow line Seyfert galaxy \pg}

Given the important implications for AGN accretion and galaxy
feedback, pioneered by \xmm\ observations, a more extended study of \pg\ was carried out over the period 2014 
June 2 to July 9, seeking to provide
greater detail on the spectral and dynamical structure of the outflow.
The resulting low background exposure of 545 ks was an order
of magnitude greater than for the any of the previous
\xmm\ observations of \pg\, in 2001, 2004 and 2007, and significantly more
than for any other UFO to date. The high statistical quality of the 2014 observation
has revealed a more complex absorption structure than previously seen in the Fe K spectrum,
providing strong evidence of multiple wind components.

Figure 3 (top panel) shows the relevant section of the stacked pn camera data plotted as a ratio to
the continuum. The emission near 6\,keV is now resolved, with component rest energies
close to the neutral Fe\,K fluorescent emission line and the 1s--2p
resonance emission lines of He-like and H-like Fe.  The
absorption line at $\sim$6.9\,keV, dominant in 2001, is now a factor $\sim$3 weaker,
with an equivalent width of
47$\pm$7\,eV. However, the much longer exposure in 2014 reveals
additional absorption structure, with significant
absorption lines at $\sim$6.6, $\sim$6.9, $\sim$7.3, $\sim$7.5, $\sim$7.8,
$\sim$8.2 and $\sim$8.7\,keV.
\footnote {The absorption lines at
$\sim$8.2 and $\sim$8.7\,keV lie close to fluorescent X-ray emission
lines of Cu and Zn arising from energetic particle impacts on the pn
camera electronics board (Strueder \et\ 2001). Fortunately, the very
low particle background throughout most of the 2014
\xmm\ observations ensured such background features have a
negligible effect on the source spectrum.}

\begin{figure}
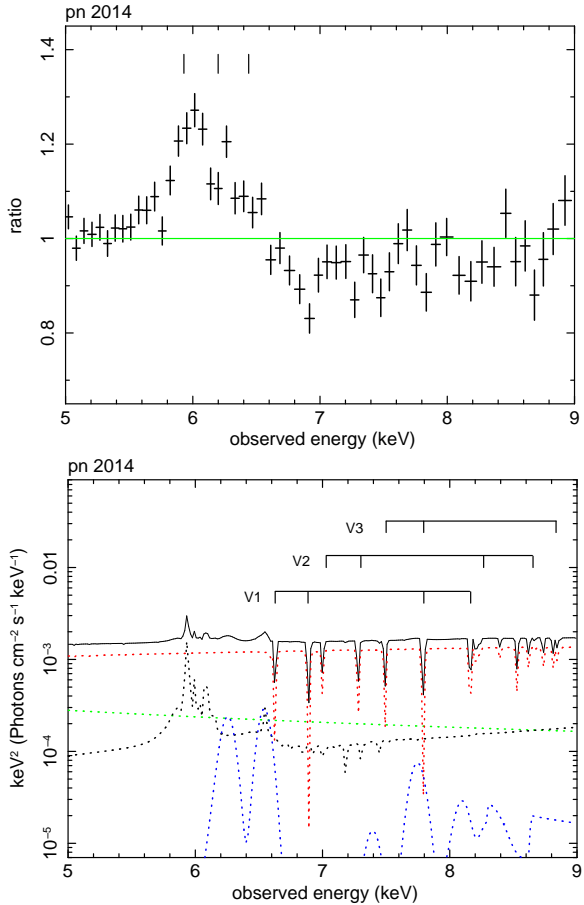
                                                          
\centering                                                              
\includegraphics[width=6cm, angle=270]{fig3_top.ps}
\centering                                                              
\includegraphics[width=6cm, angle=270]{fig3_lower.ps}       
\caption{Top: Fe\,K emission and absorption profile in the stacked
  2014 pn spectrum showing a weaker absorption line at $\sim$7\,keV compared with 2001,
  but
  additional absorption structure at higher energy and emission
  components at a small blueshift relative to markers indicating the
  rest energies of Fe\,K emission lines of neutral, He-like and H-like
  ions, respectively.  Lower: \textsc{xstar} model fit including three highly
  ionized absorption
  components. Colour coding is blue for
  the high-ionization emission, with the ionized reflection continuum and Fe\,K fluorescence line shown in
  black. The hard power law and unabsorbed power law components are in
  red and green, respectively.}
\end{figure}

While most previous UFO detections have been based on the identification of one - or at most two - blue-shifted absorption
lines, the high quality of the 2014
\pg\ data warranted more detailed modelling of the spectral features. This was done in P16 by including both
absorption and emission
spectra from ionized circumnuclear gas, modelled with grids of pre-computed spectra based on the \textsc{xstar} code (Kallman \et 1986).

Continuum reflection was also included in the P16 modelling in order to account for
an Fe\,K absorption edge that could - if ignored - result in the
line series absorption above $\sim$7\,keV being overstated.
Resolution of the Fe\,K fluorescence line, a common feature in AGN spectra,
physically linked with continuum scattering (or `reflection') from
dense matter illuminated by the nuclear hard X-ray source (Nandra
\& Pounds 1994), was used
to constrain the level of
continuum reflection. That constraint was made by adding a \textsc{xillver} component (Garcia \et\ 2013) to the spectral model, 
finding a reflection factor having only a small 
effect on the derived column densities of the absorption
components. Figure 3 (lower panel) shows a broad and shallow
ionized reflection edge in the best-fit spectral model.

Separate absorption spectra were required in P16 to match the observed spectral structure, with \textsc{xstar} parameters
corresponding to highly ionized outflows at $\sim$0.13$c$ and
$\sim$0.066$c$, comprising the dual velocity wind reported there.
The lower panel of Figure 3 shows how the principal absorption transitions in the
dual velocity absorber model (v1 and v2) match 6 of the 7 strongest
absorption lines in the stacked pn data. 
In particular, absorption lines at
$\sim$7.0, $\sim$7.3, $\sim$8.2 and $\sim$8.7\,keV correspond to $\alpha$ (resonance) and
$\beta$ transitions in Fe\,\textsc{xxv} and Fe\,\textsc{xxvi} for a velocity of 0.129$\pm$0.002$c$, while
absorption lines at  $\sim$6.6,  $\sim$6.9, $\sim$7.8 and $\sim$8.2\,keV match the same four ionic transitions,
but for the substantially lower outflow velocity of 0.066$\pm$0.003$c$.

The photoionized emission in the model is seen in Figure 3 to match
emission features in the data, the two high energy components
being identified with the He- and H-like resonance lines, in a ratio
set by the ionization parameter. The Fe Lyman-$\beta$ emission
line can also be seen in both data and model at $\sim$7.7\,keV, with a
similar blueshift.

The potential of stable pn camera data of high statistical quality to resolve and interpret such complex spectra
(notwithstanding the limited detector energy resolution: $\sim$150\,eV at 6\,keV), in combination with physically
realistic modelling,
is demonstrated by the strong preference for
a second outflow velocity in P16. The dual velocity model is
illustrated in the lower panel of Figure 3, where
the Lyman-$\alpha$ line of the lower velocity flow (v1) blends with
He-$\alpha$ from the higher velocity wind (v2) to form the observed broad
absorption feature near 7\,keV, while the He-$\alpha$ line for the lower
velocity outflow is now seen to have been partly hidden by the
emission line at $\sim$6.5\,keV.  A further
outcome of the blend at $\sim$7\,keV is to increase the relative
strength of the higher velocity Lyman-$\alpha$ line and hence the
ionization parameter and column density of that flow component.

\subsection{Evidence for a third outflow component}

Since the report of the dual velocity primary outflow was published in P16, a corresponding analysis of the soft
X-ray
spectral data from the \xmm\ RGS has been completed
(Pounds \et\ 2016b), finding counterparts of both dual velocity absorbers, as well as a higher velocity outflow
at 0.188$\pm$0.002$c$. The confirmation of 3 velocity components
in higher-resolution spectra provides strong support for the modelling of the higher-energy spectra,
while revealing the presence of
co-moving, higher density matter in each flow component, which - being less highly ionized -
retains significant opacity in the soft X-ray band.

Adding a third highly-ionized absorber to the dual velocity model of P16, with velocity v$\sim$0.19$c$, 
provides a match to the
seventh observed absorption line,  
at $\sim$7.5\,keV. The corresponding absorption series v3 in the lower panel of Figure 3 shows the Fe
Lyman-$\alpha$ and Fe He-$\alpha$ resonance absorption lines matching the observed spectral
features  at $\sim$7.5\,keV and $\sim$7.8\,keV, respectively, with the latter blending 
with the (weaker) Ly-$\beta$ line in the v2 series. In addition the Lyman-$\beta$ line of v3 now blends with the He-$\beta$
at v2 to make a better match to the strong absorption at $\sim$8.7\,keV.

\begin{table}
\centering
\caption{Parameters of the highly-ionized outflow obtained from a
2--10\,keV spectral fit to the 2014 pn data, with 3 photoionized
absorbers, defined by ionization parameter $\xi$ (erg\,cm\,s$^{-1}$),
column density $N_{\rm H}$ (cm$^{-2}$) and outflow velocity ($v/c$),
together with a photoionized emission spectrum modelled by an
ionization parameter and outflow velocity. Extracted or added
luminosities for each photoionized component are
for 2--10\,keV.}
\begin{tabular}{@{}lcccc@{}}
\hline
Comp. & log\,$\xi$ & $N_{\rm H}$($10^{23}$)  & $v/c$ & $L_{\rm abs/em}$  \\
\hline
abs 1 & 3.5$\pm$0.1 & 2.6$\pm$1.3  & 0.067$\pm$0.003 & 11$\times10^{41}$ \\
abs 2 & 3.9$\pm$0.6 & 1.5$\pm$1.0  & 0.129$\pm$0.006 & 6$\times10^{41}$ \\
abs 3 & 3.4$\pm$0.2 & 0.5$\pm$0.5  & 0.187$\pm$0.003 & 3$\times10^{41}$  \\
emission & 3.5$\pm$0.1 & 0.1(f) & 0.015$\pm$0.006 & 6$\times10^{41}$  \\
\hline
\end{tabular}
\end{table}

The simultaneous detection of 3 primary, highly ionized velocity components (Table 1)
raises a further challenge to the Eddington Winds model for the simple axisymmetric accretion disc 
envisaged in Section 2.1. We discuss in Section 4 how such further complexity might arise.

\section{Disc tearing}

In general an accretion disc around a black hole orbits in a plane
misaligned to the spin of the central black hole. This is particularly
relevant to AGN accretion where gas in the galaxy falls from far
outside the radius of gravitational influence of the central SMBH. In
this case gas is likely to fall in with essentially random
orientations (King \& Pringle 2006, 2007). When misaligned to the
black hole spin, the disc is subject to Lense-Thirring precession
which causes misaligned orbits to precess around the black hole spin
vector. This effect is strongly radially dependent with orbits at
smaller radii precessing faster. As the orbits precess, the disc
warping is communicated radially by an oscillating radial pressure
gradient around the orbit which resonates with the epicyclic motion
giving a strong torque which acts faster than the usual viscous
accretion torque (Papaloizou \& Pringle 1983; Ogilvie 1999; Nixon
2015). This effective viscosity communicating the warp through the
disc, combined with the precession from the spinning black hole
results in an aligned inner disc
connected to a misaligned outer disc by a smooth warped region
(Bardeen \& Petterson, 1975).

This picture relies on the assumption that the internal disc
communication is sufficiently fast that the disc can communicate
the precession as it is occurring. However, close to the black hole,
the precession can approach the dynamical timescale. For a black hole
spin, $a$, and black hole gravitational radius, $R_{\rm g}=GM/c^2$, the
precession frequency is $\Omega_{\rm p} = 2a (R_{\rm g}/R)^3
c^3/GM$. Thus for spins of order unity, close to the black hole the
precession can be so rapid that it is unlikely that the disc warp can
be communicated efficiently in all cases. If the disc is unable to
communicate the precession, then we expect the disc to break up into
independently precessing rings of gas. Nixon et al. (2012) derive an
approximate radius in the disc inside which this can occur (see also
Nealon et al. 2015, 2016; Dogan et al. 2015). For typical AGN disc
parameters this can be out to a few hundred gravitational radii from
the black hole. Nixon \& Salvesen (2014) discuss this model in the
context of X-ray binary state transitions. A typical tearing disc
structure is shown in Figure 4.

\begin{figure}                                                          
\includegraphics[width=8cm]{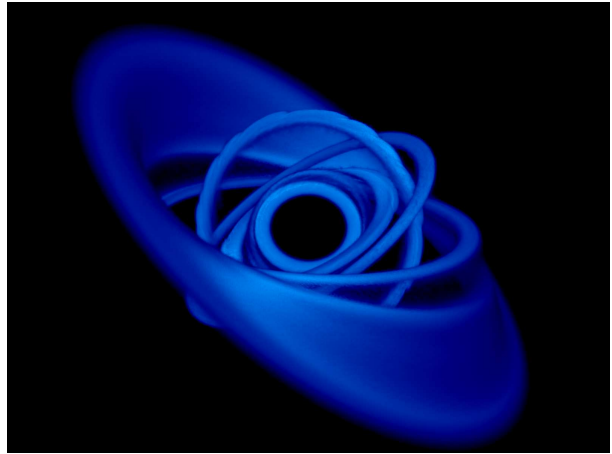}   
\caption{A disc misaligned to the central black hole spin, which has been torn apart. The innermost region is
  aligned to the
  black hole spin. Outside this the disc has been torn into discrete rings by the strong Lense-Thirring
  precession. Further
  out, the disc can communicate the precession efficiently and thus shows a smooth warp. Between precessing
  rings, the
  material shocks and loses rotational support, falling inwards to increase the instantaneous accretion rate,
  potentially
  above the local Eddington limit, thus expelling a new wind.}
\end{figure}

As the disc breaks up into discrete rings, each ring precesses on
its own timescale. After a time, two neighbouring rings precess such
that their orbital velocities become partially opposed at the contact
points. This causes shocks and internal cancellation of disc angular
momentum: while the system's total angular momentum is
conserved, the disc has borrowed angular momentum from the hole to cancel its own. The shocked material has now lost rotational
support and falls inwards to a new radius defined by its
residual angular momentum where - if it can cool fast enough - it forms a new
disc. The material added to the inner disc from the
`disc-tearing' region can result in a dramatic increase in the local accretion
rate, allowing an otherwise sub-Eddington flow to become
super-Eddington for a short time, with
the launch of winds over timescales of order days
(for radii of order $10R_{\rm g}$) to years (for radii of order
$100R_{\rm g}$), assuming a $3\times10^7M_\odot$ black hole with
near-maximal spin.

This scenario illustrates how the fast outflow properties of an AGN may be intimately linked with the
inner accretion disc physics. Detailed observations with high cadence and resolution could then provide stringent
constraints on accretion disc models and provide the opportunity to distinguish between models and refine
current understanding of these systems, which are integral to much of modern astronomy.

\section{Discussion}

The extended \xmm\ observation of \pg\ in 2014 provided uniquely
high quality hard X-ray spectra of this archetypal UFO, revealing
previously unseen structure in the $\sim$6--10\,keV energy
band. Spectral modelling has identified the observed absorption
structure with line series of H- and He-like Fe,
consistent with three distinct outflow velocities. The analysis is a strong demonstration of the
power of modelling UFOs with grids of pre-computed absorption spectra, which take proper account of all
absorption line and continua effects.

Given the success of the Black Hole Winds model (King \& Pounds 2003) in predicting the group
properties of UFOs, we consider the
more complex velocity structure in the same context.
The properties of such `Eddington winds' are described in greater
detail in King \& Pounds (2015), which includes a discussion on the
observability of UFOs as they evolve from being optically thick at
launch to transparency after radial expansion. For winds launched in
the inner accretion disc the visibility timescale for high velocity
winds can be on order months or less. That view is supported by
a review of the recent archival searches (Tombesi \et\ 2011;
Gofford \et\ 2013), leading King \& Pounds (2015) to suggest that
currently observed AGN winds are more likely to be composed of
sporadic, quasi-spherical shells, than a continuous
outflow, with only the most recently launched shells remaining
visible.

Considering the new 2014 observation of \pg\ in that context, in P16
we suggested the $v \sim$0.066$c$ and $v \sim$0.13$c$ outflows represent
recently launched shells, corresponding to local super-Eddington
episodes from prograde and retrograde accretion, with the higher efficiency of the former
(due to the inflow remaining stable to a smaller radius)
yielding the higher wind velocity (ref. equation 3). 

The simultaneous detection of 3 primary wind velocities now suggests a still more complex picture, and in Section 4 we
note that
multiple `accretion dumps' at different regions in the inner disc, leading to corresponding wind launches, might
arise from
disc warping (and breaking) due to the Lense-Thirring effect (Nixon \et\ 2012).
Monitoring the resulting pattern of disc winds over the next decade of \xmm\ observations
offers the exciting prospect of constraining detailed models of accretion and black hole growth.

While the robustness of the dual velocity absorption model described in P16, implying line resolution comparable with
the intrinsic energy resolution of the pn camera, was an indication of what can be achieved with
data of high statistical quality and from a well calibrated and stable detector, higher resolution data remain vital for observation and
interpretation of more complex outflows.
Combination of the EPIC and RGS
instruments will remain important in \xmm\ observations of AGN outflows in the years ahead.
With the unfortunate loss of the \hitomi\ spacecraft, that approach is likely to remain the best opportunity to
explore accretion physics
and black hole growth by intensive mapping of
powerful AGN available prior to the launch of \athena.

\acknowledgements
The continuing guidance of our colleague Andrew King is gratefully acknowledged
by the authors. \xmm\ is a space science mission developed and
operated by the European Space Agency and we acknowledge in particular
the excellent work of ESA staff in Madrid in successfully planning and
conducting the observations of \pg. The UK Science and Technology
Facilities Council funded the postdoctoral research assistantship of
AL and an Ernest Rutherford Fellowship for CN (ST/M005917/1).

\end{document}